\documentstyle[epsfig, psfrag,12pt]{article}
\slshape
\newcommand{\be}{\begin{equation}}
\newcommand{\ee}{\end{equation}}
\newcommand{\bef}{\begin{displaymath}}
\newcommand{\eef}{\end{displaymath}}
\newcommand{\bes}{\begin{eqnarray}}
\newcommand{\ees}{\end{eqnarray}}
\newcommand{\besf}{\begin{eqnarray*}}
\newcommand{\eesf}{\end{eqnarray*}}

\newcommand{\margen}{\hspace{8mm}}

\begin{document}

\title{Minimal String Driven Cosmology \\ and its Predictions}

\author{M.P.Infante$^{(1),(2)}$\footnote{E-mail: 
infante@mesunb.obspm.fr}, N.S\'anchez$^{(1)}$\footnote{E-mail:
Norma.Sanchez@obspm.fr} \\
{\footnotesize {\it (1) Observatoire de Paris-DEMIRM. 61, Avenue de 
l'Observatoire, 75014 Paris, FRANCE }} \\
{\footnotesize {\it (2) Dpto. F\'{\i}sica Te\'orica, Univ. de Zaragoza. 
Pza.San Francisco, 50009 Zaragoza, SPAIN}}}
\date{\empty}
\maketitle

\begin{abstract}
We present a minimal model for the Universe
evolution fully extracted from effective String Theory.
By linking this model with
a minimal but well stablished observational information, 
we prove that it gives realistic predictions on early and
current energy density and its results are compatible with 
General Relativity. Interestingly enough, the predicted current 
energy density is found $\Omega = 1$ and in anycase with lower limit
$\Omega \geq \frac{4}{9}$. On the other hand, the energy
density at the exit of 
inflationary stage gives also $\left. \Omega \right|_{inf}=1$.
This result shows an agreement with General Relativity
(spatially flat metric gives critical energy density)
within an unequivalent Non-Einstenian context (string low
energy effective equations). The order of magnitude of the
energy density-dilaton coupled term at the beginning of
radiation dominated stage agrees with GUT scale.
Without solving the known problems
about higher order corrections and graceful exit of
inflation, we find this model closer to the observational
Universe properties than the current available string cosmology 
scenarii. At a more fundamental level, this model
is by its construction close to the standard cosmological
evolution, and it is driven selfconsistently by the evolution of
the string equation of state itself. The inflationary String
Driven stage is able to reach an enough amount of inflation,
describing a Big Bang like evolution for the metric. 


{\bf PACS Numbers:} 98.80.Cq; 11.25.-w; 04.80.Cc; 98.80.-k

{\bf Report Numbers:} hep-th/9910242; DFTUZ/99/15.

\end{abstract}

\pagebreak[4]

\section{Introduction}
\setcounter{equation}0

\margen The very early stages of the Universe must be described with physics 
beyond our current models. Around the Planck time, energy and sizes 
involved would require a quantum gravity 
treatment in order to account accurately for the physics 
at such scale. 
String Theory appears as the most promising
candidate for solving the first stages evolution.
Until now, one does not dispose of a complete string theory, valid at the 
very beginning of the Universe neither 
the possibility
of extracting so many phenomenological consequences from it. 
Otherwise, effective and
selfconsistent string theories have been developed in the cosmological
context in the last years 
(\cite{tsy}-\cite{lib3}). 
These approaches 
can be considered valid at the early stages inmediately after the Planck 
epoch and should be linked with
the current stages, whose physics laws must be expected as the very low 
energy limits of the more complete laws in the early universe.
Matters raise in this process. The Brans-Dicke 
frame, emerging naturally in the low energy effective string theories, 
includes both the General Relativity as well as
the low energy effective string action as different particular cases. 
The former
one takes place when the Brans-Dicke parameter $\omega_{BD}=\infty$
while the last one requires $\omega_{BD}=-1$ (\cite{gdil}). Because
of being extracted from different gravity theories,
the effective string equations
are not equivalent to the Einstein Equations. Since current observational 
data show agreement with General Relativity predictions, whatever another 
fundamental 
theory must recover it among their lowest energy limits, or at least must 
give results compatible with those extracted in Einstein frameworks.
The great difficulties to incorporate string theory in a realistic 
cosmological framework are not so much expected at this level, but
in the description of an early Universe evolution (string phase and
inflation) compatible with the observational evolution information.
 
The scope of this paper is to present a minimal model for the Universe 
evolution completely extracted from selfconsistent string
cosmology (\cite{dvs94}). 
In the following, we recall the selfconsistent effective treatments 
in string theory and
the cosmological backgrounds arising from them. With these backgrounds,
we construct a minimal model which can be linked
with minimal observational Universe information.
We analyse the properties of this model
and confront it with General Relativity results.
Although its simplicity, interesting conclusions are found about its 
capabilities as a predictive cosmological description.
The predicted current energy density is found compatible
with current observational results, since we have $\Omega \sim 1$
and in anycase $\Omega \geq \frac{4}{9}$. The energy density-dilaton
coupled term at the beginnning of radiation dominated stage is found
compatible with the order of magnitude typical of GUT scales 
$\rho e^{\phi} \sim 10^{90} {\mathrm{erg \; cm^{-3}}}$. 
On the other hand, by defining the corresponding critical energy density, the
energy density around the exit of inflation gives $\left. \Omega \right|_{inf}
= 1$.
This result agrees with the General Relativity statement for which
$k=0 \rightarrow \Omega=1$, but it is extracted in a Non-Einstenian
context (the low energy string effective equations). No use
of observational information neither further evolution Universe
properties are needed in order to find this agreement, only
the inflationary evolution law for the scale factor, dilaton and
density energy.

Our String Driven Model
is different from previously discussed scenarii in String 
Cosmology (\cite{gv93}). Until now, no complete 
description of the scale factor evolution from inmediately post-Planckian
age until current time had been extracted in String Cosmology.
The described inflationary stage, as here presented and interpretated,
is also a new feature among the solutions given by effective
string theory.
The String Driven Model does not add new problems to the yet still open
questions, but it provides a description closer and more naturally
related to the observational Universe properties. The three stages of
evolution, inflation, radiation dominated and matter dominated are
completely driven by the evolution of the string equation of state itself.
The results extracted are fully predictive without free
parameters.

This paper is organized as follows: In Sections 2 and 3 we
construct the Minimal String Driven Model. In Section 4 we discuss its
main features: enough inflation, the evolution of the Hubble
factor and the energy density predictions. We also discuss
its main properties, differences and similarities with other
string cosmology scenarii. In Section 5 we present our Conclusions.

\section{Minimal String Driven Model}
\setcounter{equation}0

{\margen} The String Driven Cosmological Background is a minimal
model of the Universe evolution totally extracted from effective
String Theory.
We find the cosmological backgrounds from selfconsistent 
solutions of the
effective string equations.
Physical meaning of the model is preserved by
linking it with a minimal but well stablished information about 
the evolution of the observational Universe. 
Two ways allowing extraction of cosmological
backgrounds from string theory have
been used. The first one is 
the low energy effective string equations plus the string action matter.
Solutions of these equations are 
an inflationary inverse power evolution for the scale factor, as well
as a radiation dominated behaviour. On the other hand, selfconsistent 
Einstein equations plus string matter, with a classical gas of strings as 
sources, give us again a radiation dominated behaviour and a 
matter dominated description.
From both procedures, 
we obtain the evolution laws for an
inflationary stage, a radiation dominated stage and a matter
dominated stage. These behaviours are asymptotic regimes
not including strictly the transitions among stages. By
modelizing the transitions in an enoughly continuous way, we construct a 
step-by-step minimal model of evolution.
In this whole and next sections, unless opposite indication, the
metric is defined in lenght units. Thus, the (0,0) component
is always time coordinate ${\mathcal{T}}$ multiplied by constant $c$, 
$t = c {\mathcal{T}}$.
Derivatives are taken with respect to this coordinate $t$.

\subsection{The Low Energy Effective String Equations}

{\margen} We work with the low energy effective string action 
(that means, to the lowest order in expansion of powers of $\alpha'$),
which in the Brans-Dicke or string frame can be written as 
(\cite{tsy},\cite{dvs94},\cite{gv93}):
\be \label{action}
S = - \frac{c^3}{16 \pi G_D} \int d^{d+1} x \sqrt{\mid g \mid} e^{- \phi}
\left(R + \partial_{\mu} \phi \partial^{\mu} \phi - \frac{H^2}{12} + V \right)
+ S_M
\ee
where $S_M$ is the corresponding action for the matter sources, $H=dB$ is the 
antisymmetric tensor field strength and $V$ is a constant vanishing for some 
critical dimension.  The dilaton field $\phi$ depends explicitly only 
upon time coordinate and its potential will be considered vanishing. $D$ is 
the total spacetime dimension.
We consider a spatially flat background and we write the metric 
in synchronous frame ($g_{00} =1$, $g_{0i}=0=g_{0a}$) as:
\be\label{metrica}
g_{ \mu \nu}={\mathrm diag}(1, -a^2(t) \; {{\delta}_i}_j)
\ee
here $\mu, \nu = 0 \ldots (D-1)$ and $i, j = 1 \ldots (D-1)$.  
The string matter is included as a classical source which stress
energy tensor in the perfect fluid approximation takes the form:
\be \label{source}
{T_{\mu}}^{\nu} = {\mathrm diag}(\rho(t), -P(t) {{\delta}_i}^j)
\ee
where $\rho$ and $P$ are the energy density and pressure for the 
matter sources respectively.

The low energy effective string equations are obtained by extremizing 
the variation of the effective action $S$ (\ref{action})
with respect to the metric $g_{\mu \nu}$, the dilaton field
$\phi$ and the antisymmetric field $H_{\mu \nu \alpha}$ and
taking into account
the metric (\ref{metrica}) and  matter sources 
(\ref{source}). 
We will consider that the antisymmetric tensor $H_{\mu \alpha
\beta}$ as well as
the potential $B$ vanish. 
By defining $H = {{\dot{a}}\over{a}}$ and the shifted 
expressions for
the dilaton $\bar{\phi}=\phi-\ln{\sqrt{\mid g \mid}}$, matter energy
density $\bar{\rho} =\rho a^d$ and pressure $\bar{p}=P a^d$, we obtain 
the low energy effective equations  $L.E.E.$ (\cite{dvs94},\cite{gv93}):
\bes \label{leeeq}
& & {\dot{\bar{\phi}}}^{\:2} - 2 {\ddot{\bar{\phi}}} + d H^2  =  0 \\
& & {\dot{\bar{\phi}}}^{\:2} - d H^2  =  \frac{16 \pi G_D}{c^4} \; 
{\bar{\rho}} 
\; e^{\bar{\phi}} \nonumber \\
& & 2 ({\dot{H} - H {\dot{\bar{\phi}}}})  =  \frac{16 \pi G_D}{c^4} \; 
{\bar{p} 
\; e^{\bar{\phi}}} \nonumber 
\ees

The shifted expressions have the property to be invariants under the
transformations related to the scale factor duality simmetry ($a \rightarrow
a^{-1}$) and time reflection ($t \rightarrow -t$). Following this, if
($a, \phi$) is a solution of the effective equations, the dual
expression ($\hat{a}, \hat{\phi}$) obtained as:
\bes
\hat{a_i} = {a_i}^{-1} \label{dual} 
\ \ \ \ \ \ \  &  ,  &  \ \ \ \ \ \ \ 
\hat{\phi} = \phi - 2 \ln a_i 
\ees
is also a solution of the same system of equations.

\subsubsection{String Driven Inflationary Stage Solution}

{\margen} The inflationary stage appears as a new selfconsistent 
solution of the low energy effective
equations (\ref{leeeq}) sustained by a gas of stretched or unstable string
sources as developed in \cite{dvs94} and also in \cite{gv93}. 
This kind of string behaviour is characterized by a 
negative pressure and positive energy density, both growing in absolute value 
with the scale factor (\cite{dvs94},\cite{lib1},\cite{gsv2},\cite{otro}).
Strings in curved backgrounds satisfy the 
perfect fluid equation of state  $P = (\gamma - 1) \rho$ where $\gamma$ 
is different for each one of the generic three different string behaviours 
in curved spacetimes (\cite{dvs94}).
For the unstable (stretched) string 
behaviour, it holds $\gamma_u = \frac{D-2}{D-1}$ (\cite{lib1}). 
Thus, the equation of state 
for these string sources in the metric (\ref{metrica}) is given 
by (\cite{dvs94}):
\be \label{eqssd}
P = - \frac{1}{d} \ \rho
\ee

We find the following selfconsistent solution for the set of effective 
equations (\ref{leeeq})
with the matter sources eq.(\ref{eqssd}):
\bes \label{sdrin}
a(t) & = & A_I ({t_I - t})^{-Q} \ \ \ \ \ \ \ \  0 < t < t_b < t_I 
\ \ \ \ \ \ \ \   
Q = \frac{2}{d+1}  \\
\phi(t) & = & \phi_I + 2d \ln a(t)  \nonumber \\
\rho(t) & = & \rho_I {(a(t))}^{(1-d)}  \nonumber \\
P(t) & = & -{1\over{d}} \; \rho(t) \  =  \ -\frac{\rho_I}{d} 
{(a(t))}^{(1-d)} \nonumber
\ees
Notice that here $t$ is the cosmic time coordinate, running on positive 
values such
that the parameter $t_I$ is greater than the end of the string
driven inflationary regime at time $t_b$, 
$d$ is the number of expanding spatial dimensions; $\rho_I$, $\phi_I$ are
integration constants and $A_I$, $t_I$ parameters to be fixed by
the further evolution of scale factor.
 
Although the time dependence obeys a power function,
this String Driven solution is an inflationary inverse power law
proper to string cosmology. 
This solution describes an inflationary stage with accelerated expansion 
of scale factor since $H>0$, $\dot{H} > 0$ and can be considered 
superinflationary, since $\ddot{a}(t)$ increases with time. However,
notice the negative power of time and the decreasing character of
the interval $(t_I-t)$. Notice also that the string energy density
$\rho(t)$ and the pressure $P(t)$ have a decreasing behaviour 
when the scale factor grows. 

\subsubsection{String Driven Radiation Dominated Stage Solution}

{\margen} This stage is obtained by following the same procedure 
above described, but by considering now a gas of strings with dual
to unstable behaviour. Dual strings propagate in curved spacetimes
obeying a typical radiation type equation of state (\cite{dvs94},\cite{gv93}).
\be
P = \frac{1}{d} \  \rho 
\ee
This string behaviour and the dilaton ``frozen'' at constant value $(\phi =
 \mathrm{constant})$
gives us the scale factor for the radiation dominated stage:
\bes \label{sdrad}
a(t) & = & A_{II}\: t^R \ \ \ \ \ \ \ \ \ \ \ R = \frac{2}{d+1}  \\
\phi(t) & = & \phi_{II} \nonumber \\
\rho(t) & = & \rho_{II} {(a(t))}^{-(1+d)} \nonumber \\
P(t) & = & \frac{1}{d} \; \rho(t) = \frac{\rho_{II}}{d}{(a(t))}^{-(1+d)} 
\nonumber
\ees
here $\phi_{II}$, $\rho_{II}$ are integration constants, and $A_{II}$ a 
parameter to be fixed by the evolution of the scale factor. 

\subsection{Selfsustained String Universes in General Relativity}

{\margen} As shown in ref.\cite{dvs94}, \cite{lib1} and \cite{lib3},
string solutions in curved spacetimes are 
selfconsistent solutions of General Relativity equations,
in particular in a spatially flat, homogeneus and isotropic background:
\be
ds^2 = dt^2 - {a(t)}^2 dx^2 
\ee
where the Einstein equations take the form:
\bes \label{eins}
{1\over{2}}d(d-1)H^2  =   \rho \ \ \ \ \ \ \  , \ \ \ \ \ \ \
(d-1)\dot{H} + P + \rho  =  0 
\ees
As before, the matter source is described by a gas of non interacting 
classical strings (neglecting splitting 
and coalescing interactions). This gas
obeys an equation of state including the three different possible behaviours 
of strings in curved spacetimes: unstable, dual to unstable and stable. 
Let be $\mathcal{U}$, $\mathcal{D}$ and $\mathcal{S}$ the densities for 
strings with unstable, dual to unstable and stable behaviours respectively.
Taking into account the properties of each behaviour (\cite{dvs94}), the 
density energy and the pressure of the string gas are described by: 
\be \label{gasro}
\rho   =  \frac{1}{{(a(t))}^d} \left({\mathcal{U}} a(t) + {{\mathcal{D}}\over 
a(t)} + \mathcal{S}\right)  \ \ \ \ \ , \ \ \ \ \ 
P  =  \frac{1}{d} \; \frac{1}{{(a(t))}^d} \left( {{\mathcal{D}}\over a(t)} -  
{\mathcal{U}} a(t) \right) \nonumber
\ee

Equations (\ref{gasro}) are qualitatively correct for every 
$t$ and become exact in the asymptotic cases, leading to obtain the radiation 
dominated behaviour of the scale factor, as well as 
the matter dominated behaviour. 
In the limit $a(t)\rightarrow 0$ and $t \rightarrow 0$, the dual to unstable 
behaviour dominates in the equations (\ref{gasro}) and 
gives us:
\bes \label{rdrp}
\rho(t) & \sim & {\mathcal{D}} \; {(a(t))}^{-(d+1)} \ \ \ \ \ , \ \ \ \ \
P(t)  \sim  \frac{1}{d} \; {\mathcal{D}} \; {(a(t))}^{-(d+1)} 
\ees

This behaviour is characterized by positive string density energy and 
pressure, both growing when the scale factor approaches to $0$. Dual to 
unstable strings behave in similar way to massless particles, i.e. radiation. 
Solving selfconsistently the Einstein equations (\ref{eins}) with 
sources following 
eqs.(\ref{rdrp}), the scale factor solution takes the form:
\be
a(t) \sim  {\left({{2 \mathcal{D}}\over{d(d-1)}}\right)}^{1\over{d+1}} 
{\left({{d+1}\over{2}}\right)}^{2\over{d+1}} (t - t_{II})^R \ \ \ \ \ ,  \ \ 
 R = \frac{2}{d+1}
\ee
this describes the evolution of a Friedmann-Robertson-Walker radiation
dominated Universe, the time parameter $t_{II}$ will be fixed by
further evolution of the scale factor. 
On the other hand, studying the opposite limit $a(t)\rightarrow 
\infty $, $t \rightarrow \infty$ and taking into account the behaviour of 
the unstable density $\mathcal{U}$ which vanishes for
$a(t)\rightarrow \infty$ \cite{dvs94}, 
the stable behaviour becomes dominant and the 
equation of state reduces to:
\bes \label{mdrp}
\rho & \sim & {\mathcal{S}} \; {(a(t))}^{-d} 
\ \ \ \ \ \ \ \ , \ \ \ \ \ \ \ \  P \: = \: 0 
\ees 
The stable behaviour gives a constant value for the string energy, that is, 
the energy density evolves as the inverse volume decreasing with growing 
scale factor, while the pressure vanishes. Thus, stable strings behave as
cold matter. Again, from solving eqs.(\ref{eins}) with 
eqs.(\ref{mdrp}), the solution of a matter dominated stage emerges:
\be \label{rgmat}
a(t) \sim  {\left({d\over{(d-1)}}{{\mathcal{S}}\over{2}}\right)}^{1\over{d}} 
(t-t_{III})^M \ \ \  ,  \ \ \ \  M = \frac{2}{d}
\ee
We construct in the next sections a model with an inflationary stage
described by the String Driven solution (see eq.(\ref{sdrin})), followed by
a radiation
dominated stage (see eq.(\ref{sdrad})) and a matter dominated stage 
(see eq.(\ref{rgmat})). We will consider the dilaton field remain 
practically constant and vanishing from the exit of inflation until the
current time, as suggested
in the String Driven Radiation Dominated Solution.
It must be noticed that the same solution for the radiation
dominated stage emerges from the treatment with dilaton field and
without it (general relativity plus string equation of state), allowing
us to describe qualitatively the evolution of the universe by means of
these scale factor asymptotic behaviours.

\section{Scale Factor Transitions}
\setcounter{equation}0

{\margen} Taking the simplest option, we consider 
the ``real'' scale factor evolution
minimally described as:
\bes \label{real}
a_I(t) & = & A_I {(t_I - t)}^{-Q} \ \ \ \   t \in (t_i, t_r)  \\
a_{II}(t) & = & A_{II} \: t^R \ \ \ \ \ \ \ \ \  t \in (t_r, t_m) \nonumber \\
a_{III}(t) & = & A_{III} \: t^M \ \ \ \ \ \ \ \ \ t \in (t_m, t_0) \nonumber
\ees
with transitions at least not excessively long at the beginning of radiation 
dominated stage $t_r$ and of matter dominated stage $t_m$. We also define 
a beginning of inflation at $t_i$, and $t_0$ is the current time. 
It would be reasonable do not 
have instantaneous and 
continuous transitions at $t_r$ and $t_m$ for the stages extracted in the
above section, since the detail of such transitions is not provided by
the effective treatments here used. One can suspects the 
existence of very brief intermediate stages at least at the end of the
inflationary stage ($t \in (t_b,  t_r)$), as we will discuss in the next 
section, and also at the end of radiation dominated
stage ($t \sim t_m$). The dynamics of these transitions is unknown and
not easy to modelize, it introduces in anycase free parameters (\cite{p1}).
In order to construct an evolution model for the scale factor, it is
compatible with the current level of knowledge of the theory to suppose
the transitions very brief.
We will merge our lack of knowledge on the real transitions 
by means of descriptive temporal variables (\cite{p1}) in function of 
which the modelized 
transitions at $\bar{t_1}$ and $\bar{t_2}$ are instantaneous and continuous. 
We link this descriptive scale factor with the minimal information
about the evolution of the observational Universe; 
we consider the standard values for 
cosmological times: the radiation-matter transition held about 
${\mathcal{T}}_m \sim 10^{12}$ s, the beginning of radiation stage at
${\mathcal{T}}_r \sim 10^{-32}$ s  and the current age of the Universe
${\mathcal{T}}_0 \sim {H_0}^{-1} \sim 10^{17}$ s (The exact numerical
value of ${\mathcal{T}}_0$ turns out not crucial here). We impose also
to our description satisfy the same
scale factor expansion (or scale factor ratii) reached in each one of
the three stages considered in the real model (\ref{real}).
It is also convenient to fix the temporal variable of the third (and current)
stage $t$ with our physical time (multiplied by $c$). 
Explicit computations can be found in \cite{p1} and led finally to the
following scale factor in cosmic time-type variables:
\bes \label{Descr}
\bar{\bar{a_I}}(\bar{\bar{t}}) & = & \bar{\bar{A_{I}}}
{(\bar{\bar{t_I}}-\bar{\bar{t}})}^{-Q} \ \ \ \ \ \ \ 
{\bar{\bar{t_i}}}  <  {\bar{\bar{t}}} < {\bar{t_1}}   \\
\bar{a_{II}}(\bar{t}) & = & \bar{A_{II}} {(\bar{t}-\bar{t_{II}})}^R 
 \ \ \ \ \ \ \ \   \bar{t_1} < \bar{t}  < \bar{t_2}\nonumber \\
a_{III}(t) & = & A_{III} {(t)}^M 
\ \ \ \ \ \ \  \ \ \ \ \bar{t_2} <  t < {t_0} \nonumber
\ees
with continuous transitions at $\bar{t_1}$ and $\bar{t_2}$ both the 
scale factor and first derivatives with respect to the 
variables $\bar{\bar{t}}$, $\bar{t}$ and $t$.
The parameters $\bar{\bar{t_I}}$, $\bar{\bar{A_I}}$, $\bar{t_{II}}$,
$A_{III}$ can be written in function of $\bar{A_{II}}$ and 
transition times using the matching conditions.
In terms of standard observational values, the transitions $\bar{t_1}$, 
$\bar{t_2}$
and the beginning of the inflationary stage description $\bar{\bar{t_i}}$ 
are expressed as:
\bes  
\bar{t_1} & = & {R \over{M}} t_r + \left(1 - {R \over{M}}\right) t_m 
\label{T1} \ \ \ \ \ , \ \ \ \ \  \bar{t_2} = t_m  \\
\bar{\bar{t_i}} & = &
\left({R \over{M}} - {Q\over{M}}{{t_r-t_i}\over{t_I-t_r}} \right)t_r
+ \left(1 - {R \over{M}}\right) t_m \nonumber 
\ees
The parameters of the scale factor (\ref{Descr})
can be written also in terms of the observational values
$t_r$, $t_m$ and the global scale factor $\bar{A_{II}}$:
\bes \label{desc}
\bar{\bar{t_I}} & = & {t_r} {\left({R\over{M}}+{Q\over{M}}\right)} + {t_m}
{\left({1 - {R \over{M}}}\right)} \ \ \ \ , \ \ \ \ 
\bar{t_{II}}  =  \left({1-{R\over{M}}}\right) t_m  \\
\bar{\bar{A_I}} & = & \bar{A_{II}} {\left({Q \over{M}}\right)}^Q 
{\left({R \over{M}}\right)}^R {t_r}^{R+Q} \nonumber \ \ \ \ \ \ , \ \ \ \ \ \
A_{III}  =  \bar{A_{II}} {\left({R \over{M}}\right)}^{R} {t_m}^{R-M} 
\ees

The time variable $\bar{\bar{t}}$ of inflationary stage
and ${\bar{t}}$ of radiation stage are not a priori exactly equal to 
the physical time coordinate at rest frame (multiplied by $c$), but
transformations (dilatation plus translation) of it. The low energy 
effective action equations from where the scale factor, dilaton  and 
energy density have been extracted, allow these transformations.
With this treatment of the cosmological scale factor, we will attain 
computations free of "by hand" added parameters and with full predictability
as can be seen in the next sections. 

The last point is to make an approach for the dilaton field. This is
considered practically constant from the beginning of the radiation
dominated stage until the current time. 
The dilaton field increases during the inflationary string driven 
stage.  Its value can be supposed
coincident with the value at exit inflation time in a sudden but
not continuous transition, since its temporal derivatives
can not match this asymptotic behaviour.
${\phi}_{II} \: = \: {\phi} (t_r) \: \equiv \: \phi_1 $.
Remember the expression for the dilaton in inflation dominated
stage,
\be \label{dili}
{\phi}_{II} \: = \: {\phi}_I + 2 d \ln a(t_r) 
\ee

The scale factor can be written also in terms of the conformal time variable
$d\eta = \frac{dt}{a(t)}$ defined for each stage. Detailed computation
can be found elsewhere (\cite{p1}, \cite{p3}) and are not needed here.

\section{Properties of the String Driven Model}
\setcounter{equation}0

\subsection{Enough Inflation}

\margen The loose ends in standard FRW cosmology are the flatness 
problem or why the current spatial curvature $k$ is so little and the 
corresponding $\Omega$
so close to 1; the horizon problem and causality related considerations
and the large scale homogeneity of Universe and
related perturbation considerations.
The inflationary paradigm with an exponential expansion period
answers these questions provided the inflationary expansion reaches an enough
number of $e$-folders, usually considered among $80-100$.
Any other cosmological theory predicting
an inflationary stage should have to answer these questions.
As the minimum requirement in this sense, a compatibility with enough
amount of inflation must be required. 
We discuss now if
enough inflation could be provided
by the String Driven inflationary stage. 
In the comoving coordinates, enough inflation can be simply produced
if the additive parameter $t_I$ is found sufficiently near and
above the inflation-radiation dominated transition time $t_r$. 
In this way, by increasing the time, the
scale factor inflates by approaching a pole type singularity. But the 
singularity is not reached neither passed throw, because the
exit of inflation stage happens before.

Notice also that both the scale factor and dilaton
increase with time, as well as the spacetime curvature
since it is proportional to $H$. As a consequence, the
dynamics of the stage could lead to break the conditions
giving rise to the low energy effective regime. Let be
$t_b$ the instant when this breaking
in the low energy effective regime happens and, as consequence, 
the string driven inflationary evolution ends.
What kind of evolution follows is an open question, a more 
complete treatment
would be neccesary in order to describe the
scale factor in a higher order regime, where string
effects would play a strong role. By consistency with 
the time scales, it is reasonable to suppose a very short
interval between the end of the string driven inflation and
the beginning of the standard radiation dominated
cosmology. One can suspects the 
existence of a very brief intermediate stage at the end of the
inflationary stage at nearly constant curvature ($t \in (t_b, t_r)$). 
Due to its briefness and supposed bounded curvature controlled
by higher-order string corrections, not so great 
expansion would appear in the stage not governed by the $L.E.E.$. 

Let us impose the String Driven inflationary stage to
reach an arbitrary number $f$ of e-folds between the beginning
at $t_i$ and the comeback to $L.E.E.$ at $t_r$:
\be \label{e.i}
\frac{a_I(t_r)}{a_I(t_i)} \sim e^f
\ee
This condition gives us the relation:
\bef
1+\frac{t_r-t_i}{t_I-t_r} \sim e^{\frac{f}{Q}}
\eef
By defining the parameter $\epsilon$ related with the value $t_I$ through 
$t_I=t_r(1+\epsilon)$ and $\delta$ relating the inflation duration
and its exit, the enough inflation condition eq.(\ref{e.i}) translates as:
\bef
1+\frac{\delta}{\epsilon} \sim e^{\frac{f}{Q}}
\eef
that, in good approximation, gives us 
$\epsilon \sim \delta \; e^{-\frac{f}{Q}}$.

The dilaton field increases during the inflationary stage,
however its effect in breaking the $L.E.E.$ regime is not given by
the exit inflation conditions but by the beginning inflation ones.
We have modelized a time $t_i$ being the beginning of an
inflationary stage close to Planck time $t_P \sim 10^{-43} {\mathrm{s}}$. 
In this inflationary stage, from eqs.(\ref{sdrin}) the dilaton
field is $\phi(t)  =  \phi_I + 2d \ln a(t)$. Therefore, the dilaton
increases during the inflationary stage between $t_i$ and $t_r$
a ratio: 
\bef
\frac{\phi(t_r)}{\phi(t_i)} = \frac{\phi_I + 2d \ln a(t_r)}
{\phi_I + 2d \ln a(t_i)}
\eef
Thus, if an enough amount of inflation tooks place,
for instance \mbox{$a(t_r) = e^f a(t_i)$}, the ratio in the dilaton field
is given by the expression:
\bes \label{radil}
\frac{\phi(t_r)}{\phi(t_i)} & = & 1 \; + \; \frac{f}{\log_{10}e}
\frac{2d }{( \phi_I + 2d \ln a(t_i))}
\ees
where we notice the lineal dependence with the number
of e-folds $f$ and the inverse dependence of scale factor at
the beginning of inflation stage $a(t_i)$.

No need 
to place this singularity on the big bang (around $t \rightarrow 0$) is found 
from the effective equations. Still more: the beginning $t_i$ of the String 
Driven inflationary stage is practically irrelevant.
The usual inflationary paradigm
considers between 60-100 e-folds enough to
solve the problem of large scale homogeneity. This quantity is true for
standard (exponential De Sitter) inflation, but we can consider it
enoughly restrictive. In anycase, we realize that String Driven inflation
can, in principle, provide whatever amount of inflation, since a
finite value of $f$ gives a finite relation among $t_I$ and $t_r$ too. 
One would expect the value of $t_I$ given
from theory and predicting the right amount of inflation. In our limited
minimal string driven model, we prove the compatibility of our 
inflationary description with the usual standard inflation values.

\begin{figure}[h]
\centering
\psfrag{a}{$10^{-34}$}
\psfrag{b}{$10^{-33}$}
\psfrag{c}{$10^{-32}$}
\psfrag{e}[r]{$a({\mathcal{T}}_r)$}
\psfrag{f}[r]{$10^{-35}$}
\psfrag{m}[r]{$\log a({\mathcal{T}})$}
\psfrag{w}{${\mathcal{T}}$ [{\it seg}]}
\psfrag{s}[r]{{\bf SD}}
\psfrag{d}[l]{ {\bf DS}}
\psfrag{p}[r]{{\bf PL}}
\includegraphics[width=100mm]{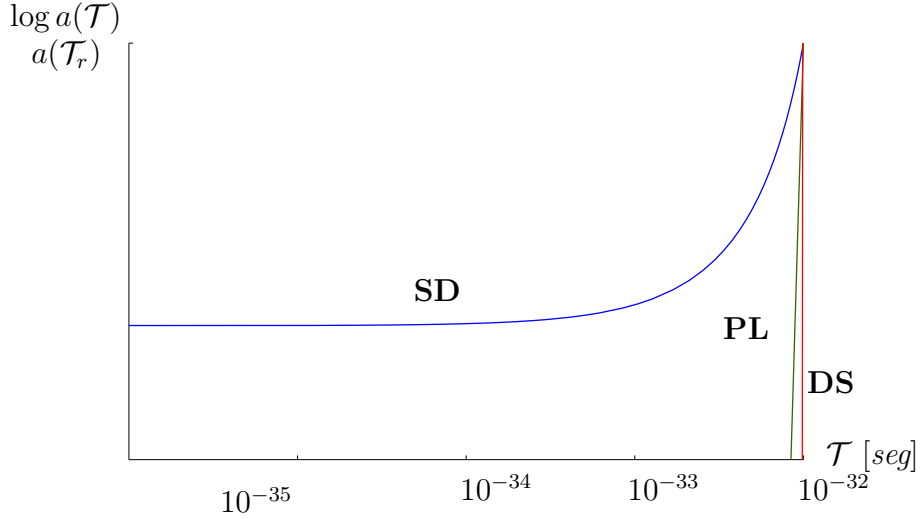}
\caption{\label{mix} Comparison among different inflationary evolution laws}
\vspace{5pt}
{\parbox{140mm}{\footnotesize Representation of the inflation amount
reached by a De Sitter $(DS)$, an usual Power Law $(PL)$ and a String 
Driven $(SD)$ inflationary stages. The three scale factors have been
normalized at ${\mathcal{T}}_r$. For enough inflation $(t_I 
\rightarrow t_r)$, the $(SD)$ behaviour becomes very much sharper.}}
\end{figure}

In the figure[\ref{mix}] we show the scale factor evolution
for a De Sitter stage($DS$), an usual power law inflationary stage ($PL$) and
the String Driven inflationary stage ($SD$).

We consider now the duration of inflation. In the descriptive coordinate
$\bar{\bar{t}}$, it means $\left. \Delta \bar{\bar{t}}\right|_{inf} = 
\bar{t_1}-\bar{\bar{t_i}}$. 
From eqs.(\ref{T1}) is seen:
\bef
\left. \Delta \bar{\bar{t}} \right|_{inf} = \frac{Q}{M}\frac{t_r}{t_I-t_r}
\left. \Delta t \right|_{inf}
\eef
where $\left. \Delta t \right|_{inf} = t_r -t_i$ and remember
$Q$ and $M$ are positive values and $t_I > t_r$. Coherently, we
obtain a positive value for the duration of inflation in descriptive
variables. 
Notice that as long as $t_I \rightarrow t_r$, this
duration experiments a greater
dilatation. In fact, if we have in general $t_I = t_r(1+\epsilon)$, then:
\bef
\left. \Delta \bar{\bar{t}} \right|_{inf} = \frac{Q}{M}\frac{1}{\epsilon}
\left. \Delta t \right|_{inf}
\eef
Since enough inflation requires $\epsilon << 1$, the
duration in descriptive variables increases so much in order
to reproduce the expansive behaviour of the scale factor. For instance,
for an expansion of order $\sim  e^f$, the last expression becames:
\bef
\left. \Delta \bar{\bar{t}} \right|_{inf} = \frac{Q}{M} e^{\frac{f}{Q}}
\left. \Delta t \right|_{inf}
\eef

Such enormous dilatation in the descriptive variables is
reached by means of transformation of the transition times. From
eqs.(\ref{T1}) we see that $\bar{t_1} \sim t_m$ and always
$\bar{t_1}>0$, meanwhile
$\bar{\bar{t_i}}$ can decrease to large negative values, enlarging
in this way the duration of the stage in the descriptive variables.
By using $t_I=t_r(1+\epsilon)$ and $t_r-t_i=\delta t_r$,
the condition $\bar{\bar{t_i}}<0$ holds when:
\be \label{tineg}
\frac{\delta}{\epsilon} \geq \frac{1}{Q} \left(R+(M-R)\frac{t_m}{t_r}\right)
\ee
With the standard transition
times given in sections above, the r.h.s. of 
eq.(\ref{tineg}) is $\sim 10^{43}$. In the l.h.s. we
find $\delta \rightarrow 1$ and $\epsilon \rightarrow 0$. For
beginning of inflation around the Planck time $(t_i \rightarrow t_P)$ and
enough inflation $(f \sim 80)$, we have 
$\frac{\delta}{\epsilon} \sim 10^{87}$ and equation (\ref{tineg})
is satisfied. This feature is needed
in order to attain the required amount of inflation in the descriptive
variables, because it requires a great dilatation of the inflation duration  
$\left. \Delta \bar{\bar{t}} \right|_{inf} \sim 10^{87} 
\left. \Delta t \right|_{inf}$.
The temporal descriptive coordinate $\bar{\bar{t}}$ runs on both
positive and negative values,  but the proper cosmic
time runs always on positive values.

The matter dominated stage description presents the
same duration that the real one since it runs on proper cosmic time.
The duration of the radiation dominated stage in descriptive
variables suffers a slight contraction with respect to duration
in proper time $\left. \Delta \bar{t} \right|_{rad} =\frac{R}{M} 
\left. \Delta t \right|_{rad}$. For our model in three spatial
dimensions, this value is $\frac{R}{M}=\frac{1}{3}$

\subsection{Evolution of the Hubble Factor}

\margen The Hubble factor in the model constructed with the String Driven
inflationary stage has the following behaviours in each stage:
\bes
\left. H(t) \right|_{inf} & = & c Q{\left(t_I-t\right)}^{-1} \\
\left. H(t) \right|_{rad} & = & c R{\left(t-t_{II}\right)}^{-1} \nonumber \\
\left. H(t) \right|_{mat} & = & c M{\left(t\right)}^{-1} \nonumber 
\ees

\begin{figure}[h]
\centering
\psfrag{t}{${\mathcal{T}}$[{\it seg}] }
\psfrag{H}{$H=c \frac{\dot{a}}{a} \; [{seg^{-1}}]$}
\psfrag{a}{$t_i$}
\psfrag{b}{${\mathcal{T}}_r$}
\psfrag{c}{${\mathcal{T}}_m$}
\psfrag{d}{${\mathcal{T}}_0$}
\psfrag{e}[r]{$H({\mathcal{T}}_0)$}
\psfrag{f}[ru]{$H({\mathcal{T}}_m)$}
\psfrag{g}[r]{$H({\mathcal{T}}_i)$}
\psfrag{h}[r]{$H_{inf}({\mathcal{T}}_r)$}
\includegraphics[width=100mm]{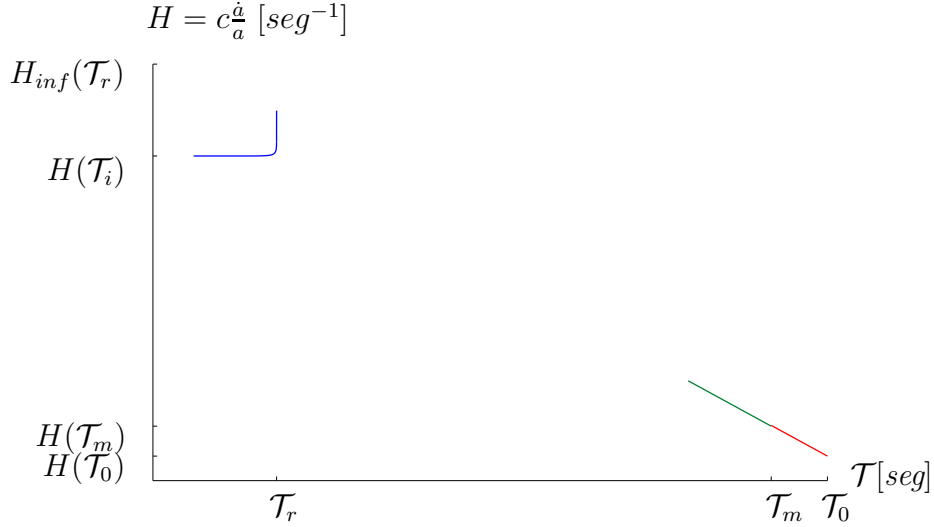}
\caption{\label{hub} Representation of Hubble factor in String Driven 
Cosmology}
\vspace{5pt}
{\parbox{140mm}{\footnotesize Representation of the Hubble factor
evolution in the minimal String Driven cos\-mo\-lo\-gi\-cal model. During the
inflationary stage, the Hubble factor remains almost constant until
the neighbourhood of the transition to radiation dominated stage. Just
before it, the Hubble factor has an explosive but bounded growth.
During the radiation dominated and matter dominated stages, the Hubble
factor decreases with a nearly constant slope.}}
\end{figure}

In the figure [\ref{hub}] we can observe the evolution of Hubble
factor in the minimal String Driven cosmological model. Notice
the growth of $H(t)$ around the exit of inflation. Since the Hubble
factor is proportional to spacetime curvature, the region around
$t_r$ is a highly curved regime, where the $L.E.E.$ regime should
transitorily break down. Notice also the nearly constant curvature
at the beginning of the inflation stage

\subsection{Energy Density Predictions}

\margen As can be easily seen, from eq.(\ref{leeeq}) we have:
\bef
{\dot{\bar{\phi}}}^{\:2} - 2 {\ddot{\bar{\phi}}} + d H^2 =  0 
\ \ \ \ \ \ , \ \ \ \ \  
{\dot{\bar{\phi}}}^{\:2} - d H^2  =  \frac{16 \pi G_D}{c^4} \; {\bar{\rho}} 
\; e^{\bar{\phi}} 
\eef
Both equations give:
\bef
\ddot{\bar{\phi}} - d H^2 = \frac{8 \pi G_D}{c^4} \; {\bar{\rho}}
 \; e^{\bar{\phi}}
\eef
By substituting the shifted expressions for the dilaton
$\bar{\phi}=\phi-\ln{\sqrt{\mid g \mid}}$ and matter energy
density $\bar{\rho} =\rho a^d$, the above equation yields:
\be \label{nume}
\ddot{\phi}-d\left(\dot{H}+{H}^2\right) \; = \; \frac{8 \pi G_D}{c^4} 
\rho \;e^{\phi}
\ee
Eq.(\ref{nume}) can be considered as the generalization of the
Einstein equation in the framework of low energy effective string action.
This equation will allow us extract some predictions on the
energy density evolution in our minimal model.

\subsubsection{Energy Density at the Exit of Inflation}
\margen By introducing the String Driven Solution for $a(t)$, $\phi(t)$,
$\rho(t)$ (eqs.(\ref{sdrin})) in eq.(\ref{nume}), we obtain a relation 
for the integration constants
$\rho_I$ and $\phi_I$:
\be \label{rdct}
\rho_I e^{\phi_I} = {{c^4}\over{8 \pi G_D}} {{2 d (d-1)}\over
{{(d+1)}^2}} {{A_I}^{-(1+d)}}
\ee
Now, is easy to find the relation between $\rho_I$, $e^{\phi_I}$ 
and the values of the energy density  $\rho_1$ and dilaton field 
$\phi_1$ at the end of inflation stage. We can compute these values 
either with the real scale factor $a_I(t)$ or with the description 
$\bar{\bar{a_I}}(\bar{\bar{t}})$. By defining
$\rho_1  =  \rho(a_I(t_r)) = \rho(\bar{\bar{a_I}}(\bar{t_1}))$ and 
$\phi_1  =  \phi(a_I(t_r)) = \phi(\bar{\bar{a_I}}(\bar{t_1}))$
and making use of eqs.(\ref{sdrin}), we can write: 
\be \label{rdin}
\rho_1 e^{\phi_1} = \rho_I e^{\phi_I} {A_I}^{1+d} 
{(t_I-t_r)}^{-Q(1+d)}
\ee

Now, with the information about the evolution of the scale factor and the
descriptions in each stage, it is possible to relate this expression 
with observational cosmological parameters. In fact, we must understand
eq.(\ref{rdin}) as one obtained in the description of inflationary 
stage:
\bef 
t_I - t_r \rightarrow \bar{\bar{t_I}}-\bar{{t_1}} =
\frac{Q}{M} t_r
\eef

For the String Driven solution, the exponents have the values $Q={2
\over{d+1}}$, $R={2\over{d+1}}$, $M={2\over{d}}$. With these 
expressions we obtain the density-dilaton coupling at the end of
inflation:
\be \label{sdri}
\rho_1 e^{\phi_1} = {{c^4}\over{8 \pi G_D}} {{2(d-1)}\over{d}} {t_r}^{-2}
\ee

With $t_r = c {\mathcal{T}}_r$ where 
${\mathcal{T}}_r \sim 10^{-32} s$, this expression gives the numerical
value:
\be
\rho_1 e^{\phi_1} = 7.1 \; \; 10^{90} \mathrm{erg} \; {\mathrm{cm}}^{-3}
\ee

It must be noticed that the same result can be achieved from the 
Radiation Dominated Stage, due to the continuity of the scale
factor, of the density energy and the dilaton field at the transition 
time $t_r$.

It must be noticed also an interesting property of this energy
density-dilaton coupled term. We can extend the General Relativity
treatment and define $\left. \Omega \right|_{inf}$
as this coupled quantity in critical energy density units, where the critical
energy density $\rho_c(t)$ for our spatially flat metric is 
$ \rho_c (t) = \frac{3 c^2}{8 \pi G} {H(t)}^2 $. We
compute the corresponding $\rho_c$ at the moment of the exit
of inflation:
\be
\left. \rho_c \right|_{\bar{t_1}} = \frac{3 c^2}{8 \pi G} H(\bar{t_1}) = 
\frac{3 c^4}{8 \pi G} M^2 {t_r}^{-2}
\ee 
With this, it is easily seen:
\be
\left. \Omega \right|_{inf} 
= \frac{\rho_1 e^{\phi_1}}{\left. \rho_c \right|_{\bar{t_1}}} 
= \frac{2}{3} \frac{d-1}{d} \frac{1}{M^2}
\ee 
where $M$ is given by eq.(\ref{rgmat}). It gives for our model in the 
three dimensional case 
$\left. \Omega \right|_{inf}=1$.  

Here we have used the descriptive variables in order to link 
the scale factor transitions with
observational transition times. But the conclusion 
$\left.\Omega \right|_{inf}=1$ is independent
from this choice. In fact, we have computed it again in the
proper cosmic time of the inflationary stage and taking
the corresponding values at the beginning of the radiation
dominated stage $t_r$. From eq.(\ref{rdin})
and $H(t) = Q {\left(t_I-t \right)}^{-1}$, we have:
\be
\left. \Omega \right|_{inf}
= \frac{1}{3Q}\left(\frac{2}{Q}-1 -dQ\right)
\ee
$Q$ is given by the String Driven solution eq.(\ref{sdrin}) and with it, 
we recover $\left. \Omega \right|_{inf} = 1$. In fact, the same result could 
be achieved also by evaluating $\left. \Omega \right|_{inf}$ 
exactly at the end of String Driven inflationary 
stage, whatever this time may be.
That means, we have a
prediction {\bf non-dependent} of whenever the exit of inflation happens.

In the proper cosmic time coordinates, $(t_I-t_r)$ is a very
little value. But independently from this, the coupled term
energy density-dilaton at the exit of inflation gives the value one in the
corresponding critical energy density units. The value
of the critical energy density is computed following the Einstein
Equations for spatially flat metrics, but solely 
the low energy effective string equations give the expression
for the coupled term eq.(\ref{nume}) and the String Driven solution
itself. In this last point, no use of further evolution, linking
with observational results neither standard cosmology have
been made. This enable us to affirm that in the $L.E.E.$
treatment, the relation between spatial curvature and 
energy density holds as in General Relativity, at least
for the spatially flat case ($k=0 \rightarrow \Omega=1$).
This result means to recover a General Relativity prescription within
a Non-Einstenian framework. 

\subsubsection{Predicted Current Values of Energy Density and Omega.}

\margen Thus, we can obtain the corresponding current value 
of $\rho_0 e^{\phi_0}$ 
in units of critical energy density or contribution to $\Omega$.

To proceed, we remember that the evolution of the density energy in
the matter dominated stage follows
$\rho \sim {a(t)}^{-3} \sim t^{-2}$
then, at the beginning of matter dominated stage we would have
\bef
\rho_m = {\left({{{\mathcal{T}}_m}\over{{\mathcal{T}}_0}}\right)}^{-2}
\rho_0
\eef
 On the other hand, in the radiation dominated stage, the density
behaviour is: $\rho \sim {a(t)}^{-4} \sim t^{-2}$
which gives for the energy density
\bef
\rho_1 = {\left({{{\mathcal{T}}_r}\over{{\mathcal{T}}_m}}\right)}^{-2}
\rho_m
\eef
\margen That is,
\bef
\rho_0 = {\left({{{\mathcal{T}}_r}\over{{\mathcal{T}}_0}}\right)}^2 \rho_1
\eef
Considering that the dilaton field has been remained almost constant
since the end of inflation $\phi_0 \sim \phi_r$, we have
\be \label{ome}
\Omega = {{\rho_0 e^{\phi_0}}\over{\rho_c}} = {\left({{{\mathcal{T}}_r}
\over{{\mathcal{T}}_0}}\right)} e^{\phi_1} {{\rho_1}\over{\rho_c}}
\ee
where the current critical energy density is expressed in terms of current
Hubble factor $H_0 = H({\mathcal{T}}_0)$ as:
\be \label{rcri}
\rho_c = {{3 c^2}\over{8 \pi G}} {H_0}^2
\ee
From eq.(\ref{ome}) and with  eqs.(\ref{sdri}) and (\ref{rcri}),   
we obtain:
\be \label{omeli}
\Omega = {{2 (d-1)}\over{3d}} {{{{\mathcal{T}}_0}^{-2}}\over{{H_0}^2}}
\ee
As like as $H_0 \sim {{\mathcal{T}}_0}^{-1}$, we have finally
\be
\Omega = {{2(d-1)}\over{3d}} 
\ee
In our three-dimensional expanding Universe, it gives $\Omega={4\over{9}}$.

 In the last, we have taken ${\mathcal{T}}_0 \leq {H_0}^{-1}$
following the usual computation. In General Relativity framework,
it holds:
\be \label{grho}
{{\mathcal{T}}_0}=\frac{2}{3}{H_0}^{-1}
\ee
if the deceleration
parameter $q_0 = -\ddot{a}(t_0)\frac{a(t_0)}{{\dot{a}(t_0)}^2}> \frac{1}{2}$.
For our model, the deceleration parameter is found:
\bef
q_0 =  \frac{1-M}{M}
\eef
that for standard matter dominated behaviour
gives exactly $q_0=\frac{1}{2}$. For this and as well as observations
give $q_0 \sim 1$, we compute the value of $\Omega$ in this 
framework. From eq.(\ref{omeli}) and eq.(\ref{grho}) we obtain:
\bef
\Omega = \frac{2}{3}\frac{d-1}{d}{\left(\frac{2}{3}\right)}^{-2}
\eef
which in the three dimensional case gives exactly:
\bef
\Omega = 1
\eef
We have obtained that a metric spatially flat $k=0$ leds to a
critical energy density $\Omega = 1$. This result is so known in
General Relativity, but here it has been extracted in a no General 
Relativity Framework, since low energy effective equations are
not a priori equivalent to General Relativity equations. Looking
from the point of view of the Brans-Dicke metric-dilaton coupling,
General Relativity is obtained as the limit of the Brans-Dicke
parameter $\omega \rightarrow \infty$, while the low
energy effective string action (see eq.(\ref{action})) is obtained
for $\omega = -1$.

Notice that $\Omega=1$ is obtained as a result of applying
coherently General Relativity statements in the matter dominated
stage, but the initial eq.(\ref{sdri}) comes for the
inflationary stage, described in the low energy effective string
framework. Thus, a result from this string treatment is
compatible with, and leads to similar predictions that,
standard cosmology.

Since in anycase ${\mathcal{T}}_0 \leq {H_0}^{-1}$, 
eq.(\ref{sdri}) can be seen as giving a lower limit for 
$\Omega$:
\be
\Omega \geq \frac{2}{3}\frac{(d-1)}{d}
\ee 
This is the prediction for the current energy density from
String Driven Cosmology and this is not
in disagreement with the current observational limits for
$\Omega$.

\subsection{Discussion of the Model}

\subsubsection{String Driven Cosmology is Selfconsistent}
\margen Matter dominated stages are not an usual result in string cosmology
backgrounds. The standard time of matter dominated beginning
${\mathcal{T}}_m \sim 10^{12}s$ is supposed too late in order
to account for string effects. But this is coherent with
the fact that matter dominated stage is described
selfconsistently by effective string treatments.

The General Relativity equations plus string sources supposes one 
of the weakest regimes for describing string-metric coupling. In
this treatment, metric evolves following classical General Relativity
equations and the string effect is accounted uniquely by considering them as
the classical matter source (\cite{dvs94}). Backreaction happens since 
the equation
of state for string behaviours have been obtained by studying their
propagation in curved backgrounds. The result is selfconsistent because
Einstein equations returns the correct curved background
when aymptotic behaviours 
have been taken. 
Thus, it is a good point that the current stage appears
selfconsistently as the asymptotic limit for large scale factor. 
Similarly, the radiation dominated stage
is obtained from both effective treatments considered.
This is coherent with having  such stage previous to
the current stage (where string effects must be not visible)
and successive to a stage extracted under considerations of stronger 
string effects. 

The intermediate behaviour between radiation and matter dominated
stages is not known. Because this and current knowledge,
this effective treatment is not able to describe suitably the
radiation dominated-matter dominated transition. As it is said
elesewhere \cite{p1}, a sudden but continuous and smooth transition among
both stages is not possible without an intermediate behaviour.
Other comment must be dedicated to the inflation-radiation
dominated transition. In the String Driven Model, this transition
requires a brief temporaly
exit of low energy effective regime for comeback within it at
the beginning of radiation dominated. This could be understood
as the conditions neccesary to modify the leading
behaviour from unstable strings to dual strings.
Further knowledge about the evolution of strings in curved
backgrounds is necessary. It has been reported that
multistring solutions (strings propagating in packets) 
are present in curved backgrounds and
show different and evolving behaviours 
(\cite{multi1}-\cite{multi3}).
Research in this sense could aid to overcome the 
transition here considered, asymptotically rounded by
low energy effective treatments.

\subsubsection{String Driven Inflation realizes a Big-Bang}

\margen During the inflationary stage, the scale factor suffers 
a great expansion, enough for solving
the cosmological puzzles. The almost amount of expansion
is reached around the exit time. In fact, the beginning of this
stage is characterized for a very, very slow evolution for
the scale factor and dilaton. This evolution increases 
speed in approaching the exit inflation, since it is
approaching also the pole singularity in scale factor.
From a phenomenological point of view, the inflationary scale
factor describes a very little and very calm Universe emerging from the
Planck scale. The evolution of this Universe is very slow
at the beginning, but the string coupling with the metric and the
string equation of state drive this evolution, leading to a each time
more increasing and fast dynamics. In approaching the exit of
inflation, the scale factor and the spacetime
curvature increase. The metric "explodes" around 
the exit of inflation. The last part of this explosion is the transition
to radiation dominated stage in a process that breaks transitorily
the effective treatment of strings. This transition is supposed
brief, the transition to standard cosmology happens at the
beginning of radiation dominated stage.

At priori, this model seems privilege an unknown parameter $t_I$, playing
the role usually assigned to singularity at $t=0$ in string cosmology.
But this is not unnatural, since in order to reach an enough amount
of inflation, this parameter is found to be very close to the 
standard radiation
dominated beginning $t_r$ and so, to the beginning of standard cosmology.
On the other hand, this value appears related to the GUT scale, 
which is consistent with 
the freezing of the dilaton
evolution and change in the equation of state in our effective
string treatment.

In this way, the Universe starts from a classical, weak coupling
and small curvature regime.  Driven by the strings, it evolves
towards a quantum regime at strong coupling and
curvature. 

 The argument above mentioned, where a brief 
transition exits and comebacks among stages
in a low energy effective treatment, 
is not an exceptional feature in String Cosmology. Pre-Big
Bang models \cite{gv93} deal with two branches 
described in low energy effective treatments. Both branches
are string duality related, but the former one runs on negative
time values. The intermediate region of high curvature
is supposed containing the singularity at $t=0$ and
consequently, the Big Bang. Our String Driven Cosmology
presents also this intermediate point of high curvature,
but it is found around the inflation-radiation dominated
transition, near but not on Planck scale.
In our model, negative proper times are never considered and the
instant $t=0$ remains before of the described inflationary stage
(is not coherent include it in the effective description, since before Planck
time a fully stringy regime is expected whose effects 
can not be considered merely with the effective equations).
There are not predicted singularities, neither at $t=0$ nor $t=t_I$, at
the level of the minimal model here studied.

Another great difference with the ``Pre-Big Bang scenario'' is
the predicted dynamics of universe. The Pre-Big Bang
scenario includes a Dilaton Driven phase running on
negative times. Not such 
feature is found
in our Universe description. Although the curvature does not
obey a monotonic regime, time runs always on positive values and
the scale factor always expands
in our String Driven Cosmology.

The Pre-Big Bang scenario assumes 
around $t \sim 0$ a ``String Phase'' with high almost constant curvature. 
Our minimal String Driven Cosmology does not
assume such a phase, but a state of
high curvature is approached (and reached) at the end of 
inflation stage. The low energy effective regime $(L.E.E.)$
breaks down around the inflation exit, both due to increasing curvature
(the scale factor approaches the pole singularity) and to the increasing 
dilaton. The exit of inflation and beginning
of radiation dominated stage must be described with a more complete
treatment for high curvature regimes.

From eq.(\ref{radil}), the growth reached by the dilaton field
during the inflationary stage would not be so large, at least
while the low energy effective treatment holds.
The exact amount depends mainly on the initial inflationary
conditions, the parameter $\phi_I$ being constrained
by the effective equations (\ref{leeeq}). In anycase, when
the scale factor approaches its singularity and the $e$-folds
number $f$ increases very quickly with time, the same is not
true for the dilaton ratio. Comparatively, the dilaton ratio increases in
a much slower way $\frac{\phi(t_r)}{\phi(t_i)} \sim f$ than the scale
factor $\frac{a(t_r)}{a(t_i)} \sim e^f$. As a consequence, 
corrections due to the high curvature regime are needed
much earlier than the corresponding to dilaton growth.

\section{Conclusions}
\setcounter{equation}0

\margen We have considered a minimal model for the evolution of the 
scale factor
totally extracted from string cosmology. 
The earliest stages (an inflationary  power type expansion and a radiation 
dominated stage) are obtained from the low energy effective equations, 
while the radiation dominated
stage and matter dominated stage
are selfconsistent solutions of the Einstein equations with
a gas of strings as matter sources. 
Such solutions suggest the low energy
effective action is asymptotically valid at earliest stages, around
and immediately after Planck time $t_P \sim 10^{-43} {\mathrm{s}}$, when
the spacetime and string dynamics would be strongly coupled.
The radiation dominated stage is extracted
from both treatments, coherently with being it an intermediate
stage among the two regimes: inflation and matter dominated stage. 
On the other hand, since the stable string 
behaviour
describes cold matter behaviour, current matter dominated stage
can be extracted also in the string matter treatment.
Notice that in 
string theory, the equation of state of the string matter is derived 
from the string dynamics itself and not given at hand from outside
as in pure General Relativity.

No detail on the transitions dynamics can be extracted in this
framework, too naive for accounting such effects. The 
inflation-radiation dominated transition implies a transitory breaking of the 
low energy effective regime. The radiation dominated-matter dominated stage
can not be modelized in sudden, continuous and smooth way.
Further study is required on the evolution of the equation of state for 
the gas of
strings. The three string behaviours
in curved backgrounds are present in this gas and each cosmological 
stage is selfconsistently driven by them.
In this way, the transition
between inflation and radiation dominated stage is related with the 
evolution from unstable to dual string behaviour, while the
radiation dominated-matter dominated transition could be driven by passing
from the dual to stable behaviour.

Phenomenological information extracted from this String Driven
model is compatible with observational Universe information.
An amount of inflation, usually considered as enough for solving
the cosmological puzzles, can be obtained in the inflationary stage.
Energy ranges at exit of inflation are found coherents with GUT scales.
The inflationary stage gives a value for the energy
density-dilaton coupled term equivalent to the corresponding
critical energy density, computed as in General Relativity. That
means, we have $\left. \Omega \right|_{inf} = 1 $, whenever the
end of inflationary stage be computed. Also, the
contribution to current energy density is computed $\Omega \geq
\frac{4}{9}$, and taking account the validity of General
Relativity in the current matter dominated stage,
we find this contribution be exactly $\Omega = 1$.

Our main conclusion is to have proved that string cosmology, 
althought being effective, is able to produce a
suitable minimal model for Universe evolution.
It is possible to place each effective context in a time-energy
scale range. Energy ranges are
found and General Relativity conclusions are obtained
too in a string theory context coherently. We have extracted the 
General Relativity statement about spatial curvature and 
energy density, at least for the spatially flat case 
($k=0 \rightarrow \Omega=1$) in a totally Non-Einstenian
framework, like the low energy effective string action giving
rise to the inflationary String Driven stage.  

In their validity range,  no need of extra stages
is found. Only the interval around the transitions and the
very beginning epoch, probably the Planck epoch, will require
more accurate treatments that hitohere
considered.
Since the behaviours above extracted are asymptotic results, it is
not possible to give the detail of the transitions among the different 
stages. 
The connection among asymptotic low energy effective regimes through
a very brief stage (requiring a more complete description of
string dynamics) enables us to suppose this stage containing the evolution
in the equation of state from unstable strings domain to dual strings one.
From the point of view of the scale factor evolution, this brief transition
could be modelized as nearly instantaneous, provided curvature and
scale factor expansion have attainted nearly their maximun values.
Similar comment can be made in the radiation dominated-matter 
dominated transition. It must be driven by the subsequent evolution of 
the gas of strings from dual to unstable domain to string stable behaviour
domain.
Again, a brief intermediate stage could take place
among both asymptotic behaviours. 
But this is an open 
question in the framework of string cosmology both for inflation-radiation 
dominated as well as radiation dominated-matter dominated transition.

{\bf ACKNOWLEDGMENTS}

M.P.I. wants to thank Marina Ram\'{o}n Medrano and H\'{e}ctor
J. de Vega for very helpful discussions.

\end{document}